\documentstyle[12pt]{article}

\catcode`\@=11

\def\@normalsize{\@setsize\normalsize{15pt}\xiipt\@xiipt
\abovedisplayskip 14pt plus3pt minus3pt%
\belowdisplayskip \abovedisplayskip
\abovedisplayshortskip  \z@ plus3pt%
\belowdisplayshortskip  7pt plus3.5pt minus0pt}

\def\small{\@setsize\small{13.6pt}\xipt\@xipt
\abovedisplayskip 13pt plus3pt minus3pt%
\belowdisplayskip \abovedisplayskip
\abovedisplayshortskip  \z@ plus3pt%
\belowdisplayshortskip  7pt plus3.5pt minus0pt
\def\@listi{\parsep 4.5pt plus 2pt minus 1pt
            \itemsep \parsep
            \topsep 9pt plus 3pt minus 3pt}}

\def\underline#1{\relax\ifmmode\@@underline#1\else
        $\@@underline{\hbox{#1}}$\relax\fi}
\@twosidetrue

\relax

\catcode`@=12

\catcode`\@=11

\def\ps@headings{\def\@oddfoot{}\def\@evenfoot{}
\def\@oddhead{\hbox{}\hfill
        \makebox[.5\textwidth]{\raggedright\ignorespaces --\thepage{}--
\hfill {}}}}

\def\@evenhead{\@oddhead}
\ps@headings

\catcode`\@=12

\relax

\def\figcap{\section*{Figure Captions\markboth
        {FIGURECAPTIONS}{FIGURECAPTIONS}}\list
        {Fig. \arabic{enumi}:\hfill}{\settowidth\labelwidth{Fig. 999:}
        \leftmargin\labelwidth
        \advance\leftmargin\labelsep\usecounter{enumi}}}
 \relax
\def\tablecap{\section*{Table Captions\markboth
        {TABLECAPTIONS}{TABLECAPTIONS}}\list
        {Table \arabic{enumi}:\hfill}{\settowidth\labelwidth{Table 999:}
        \leftmargin\labelwidth
        \advance\leftmargin\labelsep\usecounter{enumi}}}
 \relax
\def\reflist{\section*{References\markboth
        {REFLIST}{REFLIST}}\list
        {[\arabic{enumi}]\hfill}{\settowidth\labelwidth{[999]}
        \leftmargin\labelwidth
        \advance\leftmargin\labelsep\usecounter{enumi}}}
 \relax

\catcode`\@=11

\def\@evenhead{\@oddhead}

\ps@headings

\relax

\newskip\humongous \humongous=0pt plus 1000pt minus 1000pt

\newif\ifdtup

\def\tr{\mathop{\rm tr}}

\def\Im{\mathop{\rm Im}}
\def\Re{\mathop{\rm Re}}

\def\abs#1{\left| #1\right|}

\def\beq{\begin{equation}}
\def\eeq{\end{equation}}

\def\beqn{\begin{eqnarray}}
\def\eeqn{\end{eqnarray}}
\relax

\def\G2{{\; \rm GeV/}c^2}
\def\G{\; \rm GeV}

\def\dotx{\dotx{\dot\overline{x}}}

\relax

\def\tcdot{\cdot\cdot\cdot}

\begin{document}

\begin{titlepage}
\nopagebreak
\begin{flushright}

        {\normalsize    OU-HET 188  \\
                           April,~1994  \\}

\end{flushright}

\vfill
\begin{center}
  {\large \bf         Multicritical Behavior of \\
                          $c=1$ Matrix Model
                       }\footnote{This work is supported in part by
  Grant-in-Aid for  Scientific Research
  $(05640347)$  from the Ministry of Education, Japan.}

\vfill

         {\bf H.~Itoyama} ~~ and ~~         {\bf M.~Koike}

        Department of Physics,\\
        Faculty of Science, Osaka University,\\
        Toyonaka, Osaka, 560 Japan\\

\end{center}
\vfill

\begin{abstract}
   We discuss multicritical behavior of $c=1$ matrix model, extending
   the recent work of ref. \cite{CIO} on  a nonperturbative completion
  of the density of states function.
 For the odd orders of multicriticality,   we are able to determine
 the higher genus contributions and a nonperturbative completion
 from the WKB wave function of the multicritical periodic potential.  The
 expression for the contributions as a function of the scaled chemical
 potential is found to be the same as  the one at the lowest critical point.
  We point out a strange scaling behavior.

\end{abstract}
\vfill
\end{titlepage}

 An  innumerable amount of work has been done in matrix models describing
 discretized string models with $c \leq 1$ coupled to
 two-dimensional gravity\cite{BKDSGM,Kazakov,c=1}.
  Relatively little attention has been paid, however, to the behavior of
 $c=1$ models at their multicritical points.
   There are definitely some issues which deserve further study
 at these multicritical points  which may clarify the role of discrete
 states, properties of the $c=1$ barrier and others.
  In this letter, we  present our calculation of the higher genus
 contributions to the density of states function
   at the odd orders of multicriticality which follow from  a nonperturbative
 completion due to the periodic potential.
 We give a brief physical discussion in the end of the paper.

  Our calculation is  regarded as an application of  the recent work
 by Chaudhuri, Ooshita and  the one of the authors \cite{CIO}.
 In \cite{CIO}, a novel way of evaluating nonperturbative contributions
 has been given.   The conventional computation of $c=1$ matrix
 models starts from  the quantum mechanics of the inverted harmonic
 oscillator \cite{c=1}. The potential is bottomless and this fact makes it
 difficult to go beyond  the perturbative evaluation. The work of \cite{CIO}
  starts instead from the quantum mechanics of
 the well-defined periodic potential which follows from the one-plaqutte
  hamiltonian of $U(N)$ lattice gauge theory \cite{GrossWitten,Wadia,Neu}.
  The expression for the density
 of states through the dispersive integral
 is shown to be finite and the asymptotic (perturbative) series can be
 recovered  by blowing up the region of the singularity in the exact
 expression. This is, therefore, a way of making  a nonperturbative completion
 of the asymptotic series\footnote{ The answer obtained
 in ref.\cite{CIO} will be one among  infinitely many ways
 of making a nonperturbative completion.  The perspective
 is that this   would give us  a more constructive
  procedure to address the issue of  nonperturbative universality.}.
One can turn the argument around. The procedure of \cite{CIO} provides
 an efficient way of producing a perturbative series  from a potential
  with a bottom and this is what we apply to the multicritical points
 of the $c=1$ matrix model.

  The quantum mechanics hamiltonian we use in what follows is
\beq
\label{eq:h0}
\hat{h}^{(n)}_0=-\frac{2\lambda}{N^2}\frac{d^2}{d\theta^2} +
 \frac{2^{n+1}}{\lambda} -\frac{2}{\lambda}\left(1+\cos\theta\right)^n\;\;\;.
\eeq
  We denote its eigenvalue by $\epsilon$.    Alternatively, we often write
\beq
\label{eq:h1}
\hat{h}^{(n)}=-\frac{2\lambda}{N^2}\frac{d^2}{d\chi^2} -
\frac{2}{\lambda}\left(1-\cos\chi\right)^n \;\;\;
\eeq
  whose eigenvalue is denoted by $\mu$.
  Here  $\chi\equiv\pi - \theta$ and
 $\mu = \epsilon - \frac{2^{n+1}}{\lambda}$.
  The hamiltonian (eqs.~(\ref{eq:h0}),(\ref{eq:h1})) introduced above
can be thought of as the one
 following from the one-plaquette hamiltonian  of the form
\beq
H=-\left(\frac{g^2}{2}\right)\sum_{{\scriptstyle \alpha}
\atop{\scriptstyle{\rm all\ link}}}
 E_{\ell,\alpha} E_{\ell,\alpha} + \frac{1}{g^2}
 \tr\sum_n \left(g_n U^n + g^\ast_n U^{\dagger n}\right)\;\;\;
\eeq
 when acting on the gauge invariant subspace.
 We leave the notation to  \cite{CIO,Wadia} as details are not important
 here.

  Let us now begin with determining the singularity of the ground state energy
  in the planar limit.  A brief discussion is seen  in \cite{GZ} and we present
 a minimum account here.  The density of states function in the planar limit is
\beqn
\rho^{(n)}_{pl}&\equiv&\lim_{N\rightarrow\infty}\rho_N^{(n)}\nonumber\\
  &\equiv& \int_{0}^{2 \pi} \frac{d \theta}{2\pi} \int_{-\infty}^{\infty} dp
  \delta \left( \epsilon - 2\lambda p^{2} - \frac{2^{n+1}}{\lambda}
  + \frac{\lambda}{2}(1+\cos \theta)^{n} \right)  \nonumber \\
&=& \frac{1}{\sqrt{2^n}\pi}\int_0^{\pi/2}d\theta\frac{\Theta\left(k
 + \cos^{2n} \theta\right)}{\sqrt{k + \cos^{2n} \theta}} \;\;\;,
\eeqn
  where $k\equiv \epsilon\lambda/2^{n+1} -1=
   \mu\lambda/2^{n+1}$.
  For $ k>0 $,
\beqn
\label{eq:stpl}

\rho^{(n)}_{pl}\left(\epsilon,\lambda\right)=\frac{1}{\sqrt{2^n}\pi}\int_0^{\pi/2}d\theta\frac{1}{\sqrt{k+\cos^{2n} \theta}}
=\frac{1}{2\sqrt{2^n} \pi}\int_0^1 dt
\frac{1}{\sqrt{t\left(1-t\right)\left(k+t^n\right)}}\;\;,
\eeqn
  while, for $k<0 $,
\beqn
 \rho^{(n)}_{pl}\left(\epsilon,\lambda\right)
=\frac{1}{\sqrt{2^n}\pi}\int_0^{\cos^{-1}\left(\abs{k}^{1/n}\right)}
 d\theta\frac{1}{\sqrt{k+\cos^{2n} \theta}}\;\;.
\eeqn
  Eq.~(\ref{eq:stpl}) is the Riemann-Liouville integral of order
 $1/2$ \cite{RL}
 and is expressible in terms of a generalized hypergeometric function
\beq
\rho^{(n)}_{pl}\left(\epsilon,\lambda\right) =
\frac{1}{2\sqrt{2^n} \pi}\frac{\Gamma\left(1/2\right)^{2}}
{\sqrt{k}}\, {}_{n+1}F_n \left(\frac{1}{2},\frac{1}{2n},
\frac{3}{2n},\tcdot,\frac{2n-1}{2n};\frac{1}{n},\frac{2}{n},
\tcdot,1;-\frac{1}{k}\right) \;\;.
\eeq
 Using the asymptotic formula \cite{RL}
\beq
{}_{n+1} F_n \left(  \cdots
;-\frac{1}{k}\right)\sim\frac{\Gamma\left(\frac{1}{2}-\frac{1}{2n}\right)\Gamma\left(\frac{1}{2n}\right)}{\Gamma\left(\frac{1}{2}\right)}\prod_{j=1}^n\left(\frac{\Gamma\left(\frac{j}{n}\right)}{\Gamma\left(\frac{2j-1}{2n}\right)}\right)^2 k^{\frac{1}{2n}} \;\;,
\eeq
  we find, as $k \rightarrow 0 $,
\beqn
\label{eq:rhop}
\rho^{(n)}_{pl}\left(\epsilon,\lambda\right)
&=&\frac{1}{2\sqrt{2^n}n \pi^{3/2}}
\Gamma\left(\frac{1}{2}-\frac{1}{2n}\right)\Gamma\left(\frac{1}{2n}\right)
k^{-\left(\frac{1}{2}-\frac{1}{2n}\right)}  \;, \;\;   n= 2,3, \cdots .
\eeqn
 In order to express the ground state energy  as a function of the
  cosmological constant, we invert the normalization condition
\beqn
1=\int_{-2^{n+1}/\lambda }^\mu d\mu'\rho^{(n)}_{pl}
  \left(\mu',\lambda\right)
 =\frac{\lambda_c}{\lambda} \frac{\Gamma\left(\frac{n}{2}+1\right)}
{\Gamma\left(\frac{n+1}{2}\right)\Gamma\left(\frac{1}{2}\right)}
\int_0^1 dt     \sqrt{\frac{k+t^n}{t\left(1-t\right)}} \;\;\;,
\eeqn
  where
$\lambda_c \equiv\frac{\Gamma\left(\frac{n+1}{2}\right)
\Gamma\left(\frac{1}{2}\right)}{\Gamma\left(\frac{n}{2}+1\right)}
\frac{2\sqrt{2^n}}{\pi} $ and  $x\equiv\frac{\lambda -\lambda_c}{\lambda_c}$
 is the renormalized cosmological constant $t$ multiplied by the square
 of an auxiliary parameter $a$ (lattice spacing).
  We find
\beq
\label{eq:kx}
k=\left[\pi\left(n+1\right)
 \frac{\Gamma\left(\frac{n+1}{2}\right)}{\Gamma\left(\frac{n}{2}+1\right)}
\frac{1}{\Gamma\left(\frac{1}{2}-\frac{1}{2n}\right)\Gamma
\left(\frac{1}{2n}\right)}x\right]^{\frac{2n}{n+1}} \;\;\;.
\eeq
  After some calculation with partial integration and
 the subtraction of a nonuniversal constant,
 we obtain the planar ground state energy
\beqn
  \lim_{N\rightarrow \infty}  \frac{1}{N^{2}} E^{gr}
 = \int_{-\frac{2^{n+1}}{\lambda}}^\mu d\mu'
 \epsilon' \rho^{(n)}\left(\mu';\lambda\right)
 \sim \frac{2^{\frac{3}{2}n+2}}{\lambda_c^2}\frac{1}{2\sqrt{\pi}}
\frac{2n}{3n+1}  \;\;\; \nonumber \\
 \times \Gamma\left(\frac{1}{2}-\frac{1}{2n}\right)
\Gamma\left(\frac{1}{2n}\right)\frac{1}{n\pi}\left[\pi\left(n+1\right)
\frac{\Gamma\left(\frac{n+1}{2}\right)}{\Gamma\left(\frac{n}{2}+1\right)}
\frac{1}{\Gamma\left(\frac{1}{2}-\frac{1}{2n}\right)\Gamma
\left(\frac{1}{2n}\right)}x\right]^{2+\frac{n-1}{n+1}}
\eeqn
 up to the leading singularity.
  The string susceptibility is
\beq
  \gamma_{str} = - (1- \frac{2}{n+1}) \;\;\;\;\;
\eeq
  confirming \cite{GZ}.

  We now proceed  to  the nonperturbative evaluation of the density of states
  by taking the double scaling limit. Let $ \psi^{(\pm)}$ be a set of
   eigenfunctions whose semi-classical form  is given by
\beqn
\psi^{\left(\pm\right)}_{WKB}=\frac{1}{\sqrt{\tilde{p}^{(n)}}}
\exp\left[\pm iN\int_\pi^\chi \tilde{p}^{(n)} \left(\chi' \right)d\chi' \right]
,{\rm with}\;
\tilde{p}^{(n)} \equiv\sqrt{\frac{1}{2\lambda}}
\sqrt{\frac{2^{n+1}}{\lambda}\sin^{2n}\frac{\chi}{2} + \mu} \;.
\eeqn
   By letting $N$ sufficiently large, we can make the semi-classical
 approximation valid except at the turning point.

  The integral $\int_\pi^\chi d\chi' \tilde{p}_n ( \chi^{\prime})$
 can be carried out term by term  as a power series in  $k$.
  The answer   for  $n=$ odd is qualitatively different  from the one for
 $n=$ even.  For $n=$ odd,  the answer contains logarithms as is in the case
 of the lowest critical point \cite{CIO}, so that the formalism of
  \cite{CIO} is applicable.   For $n=$ even, the answer contains a term
 proportional to $ \chi -\pi$. We have not been able to understand
 this difference in physical terms. From now on, we restrict ourselves to
 the $n=$ odd cases.  The integral is expressible as
\beqn
\int_\pi^\chi d\chi' \tilde{p}^{(n)} ( \chi^{\prime})
 &=&
R\left(\frac{\chi}{2}\right)-\frac{\sqrt{2^n}}{\lambda}g\left(-k\right)\ln\tan\frac{\chi}{4} \;\;, \;  \;\;\;  \\
    {\rm where}\;\;
g\left(x\right) &=&   \sum_{i=1}^\infty \frac{1}{i!}
\frac{\left(2i-3\right)!!}{2^{i-1}}
\frac{\left[\left(2i-1\right)n-2\right]!!}
{\left[\left(2i-1\right)n-1\right]!!}x^i  \;\;\; \nonumber \\
 &=&\frac{\left(n-2\right)!!}{\left(n-1\right)!!}x\, {}_{n+1}F_n
 \left(\frac{1}{2},\frac{1}{2},\frac{1}{2}+\frac{1}{n},\tcdot,\frac{3}{2}
-\frac{1}{n}\right.\cr
& &\vspace{30pt}\left.;2,\frac{1}{2}+\frac{1}{2n},
\frac{1}{2}+\frac{3}{2n},\tcdot,\frac{1}{2}+\frac{n-2}{2n},\frac{1}{2}
+\frac{n+2}{2n},\tcdot,\frac{3}{2}-\frac{1}{2n};x\right)
\eeqn
   and
\beqn
R\left(\frac{\chi}{2}\right)= \frac{\sqrt{2^n}}{\lambda}
\left[-2\frac{\left(n-1\right)!!}{n!!}\sum_{j=0}^{\frac{n-1}{2}}
\frac{\left(2j-1\right)!!}{\left(2j\right)}
\cos\frac{\chi}{2}\sin^{2j}\frac{\chi}{2}\right.\nonumber\\
\left. +\sum_{i=1}^\infty \frac{1}{i!} \frac{\left(2i-3\right)!!}{2}
\frac{\left[\left(2i-1\right)n-2\right]!!}{\left[\left(2i-1\right)n-1\right]!!}
 \sum_{j=1}^{\frac{\left(2i-1\right)n-1}{2}}
\frac{\left(2j-1\right)!!}{\left(2j\right)!!}
 \frac{2\cos\frac{\chi}{2}}{\sin^{2j}\frac{\chi}{2}}
 \left(-k\right)^i \right] \;\;\;.
\eeqn
 Function $R$ satisfies
\beq
R\left(-\frac{\chi}{2}\right)=R\left(\frac{\chi}{2}\right),\ R
\left(\frac{\chi+2\pi}{2}\right)=-R\left(\frac{\chi}{2}\right) \;\;.
\eeq

 The actual construction of  wave functions
 as well as the formula for the density of states through the dispersion
 relation can be given exactly as in \cite{CIO}. We leave the details to
 \cite{CIO}.
\beqn
\psi^{\left(R\right)}_\pm \left(\chi ;\mu\right)&=&A\psi^{\left(+\right)}
\left(\chi ;\mu\right)+A^\ast
 \psi^{\left(-\right)}\left(\chi ;\mu\right) \nonumber\\
&\pm& \left(B\psi^{\left(+\right)}
\left(\left\vert\chi_r \right\vert ;\mu\right) +
 B^\ast \psi^{\left(-\right)} \left(\left\vert\chi_r \right\vert
 ;\mu\right)\right)\hspace{30pt} 0<\chi <\pi   \nonumber \\
\psi^{\left(L\right)}_\pm \left(\chi ;\mu\right)&=&
B\psi^{\left(+\right)}\left(\left\vert\chi
 \right\vert ;\mu\right) + B^\ast \psi^{\left(-\right)}
 \left(\left\vert\chi \right\vert ;\mu\right)\nonumber\\
&\pm& \left( A\psi^{\left(+\right)}
\left(\chi_r ;\mu\right)+A^\ast \psi^{\left(-\right)}
\left(\chi_r ;\mu\right) \right) \hspace{40pt} -\pi <\chi <0
\eeqn
is the form of the wave function
 which is  continuous over   the entire domain , periodic and real
 and possess a definite parity.
  The coefficients $A$ and $B$ are related through the matching condition
  betwween the wave function at domain $(L)$ and  the one at $(R)$:
\beq
A=e^{-\pi NG\left(\mu,\lambda\right)} B\;\;  {\rm with}~
G\left(\mu,\lambda\right)=-\frac{\sqrt{2^n}}{\lambda}
g\left(-k\right) \;\;.
\eeq
    The normalization  condition over an entire period gives
\beqn
\label{eq:abs}
\abs{A}^2 \approx \frac{1}{4}\frac{e^{\pi NG}}{\cosh \pi NG}
\left(\int^\pi \frac{d\chi}{\tilde{p}\left(\chi\right)}\right)^{-1}
 \;\;, \;\;\;
\abs{B}^2 \approx \frac{1}{4}\frac{e^{-\pi NG}}{\cosh \pi NG}
\left(\int^\pi \frac{d\chi}{\tilde{p}\left(\chi\right)}\right)^{-1} \;\;\;.
\eeqn
  The density of states is given by
\beq
\rho_N \left(\mu\right)=\frac{1}{2\pi N}\frac{\partial}{\partial \mu}
 \left(\arg A + \arg B\right),
\eeq
 and this can be computed from eq.~(\ref{eq:abs}) by the dispersion relation.
  A nontrivial scaling limit is obtained  by letting $N$ to infinity
$ {\displaystyle  \rho\left(\mu\right)= \lim_{N\rightarrow\infty}
 \rho_N \left(\mu\right)  }$:
\beqn
\label{eq:rho}
  \rho \left( \mu \right)&=&-\frac{\lambda}{4\pi}
\lim_{N\rightarrow\infty} \lim_{L\rightarrow\infty}
 \Im \left(\sum_{\ell=0}^L \frac{1}{Niy_{\ell} +\tilde{k}^2}
\right)\nonumber\\
&\equiv&-\frac{\lambda}{4\pi}
 \lim_{N\rightarrow\infty} \lim_{L\rightarrow\infty} q \left(N,L\right) \;\;\;,
\eeqn
 where $ \tilde{k}^{2} = -Nk$ and
\beqn
q\left(N,L\right)&=&\sum_{\ell=0}^L \frac{N\Im g^{-1}
\left(i\frac{\lambda}{\sqrt{2^n}}\frac{\ell+1/2}{N}\right)}
 {\left(N\Re g^{-1} \left(i\frac{\lambda}{\sqrt{2^n}}
\frac{\ell+1/2}{N}\right) - \tilde{k}^2 \right)^2 +
 \left(N\Im g^{-1}\left(i\frac{\lambda}{\sqrt{2^n}}\frac{\ell+1/2}{N}\right)
\right)^2} \nonumber\\
&\equiv& -\sum_{\ell=0}^L a_{\ell}\left(N\right) \;\;\;.
\eeqn
  Following \cite{CIO}, one can show that the infinite series
 $  {\displaystyle \lim_{ L \rightarrow \infty}
 \sum_{\ell = 0}^{L} a_{\ell}(N) } $ is
 convergent and bounded by a number independent of $N$.  This means that
 $\rho(\mu)$ is finite.

  Having obtained a nonperturbative  formula for the density of states, we
 proceed to find the asymptotic expansion $\rho_{\rm as}\left(\mu\right)$
  by interchanging the two limits
 $N\rightarrow\infty$  $L\rightarrow\infty$  in eq.~(\ref{eq:rho}) \cite{CIO}.
   The function $g^{-1}(x)$ gets linearlized in this limit:
 $g^{-1}\left(x\right)\rightarrow \left(n-1\right)!!/\left(n-2\right)!!$.
  We find
\beqn
\label{eq:rhoas}
\rho_{\rm as}\left(\mu\right) &\equiv& \frac{\lambda}{4\pi}
\lim_{L\rightarrow\infty} \lim_{N\rightarrow\infty} \Im \left(\sum_{\ell=0}^L
\frac{1}{iNy_{\ell} -\tilde{k}^2}\right)\nonumber\\
&=& \frac{\sqrt{2^n}}{4\pi}\frac{\left(n-1\right)!!}{\left(n-2\right)!!}
\left[\left(\mbox{planar term}\right) + \sum_{m=1}^\infty \left(2^{2m-1}-1
\right)\frac{\left\vert
B_{2m}\right\vert}{m}\frac{1}{\tilde{\xi}^{2m}}\right]\;\;\;,
\eeqn
where $\tilde{\mu}=N\mu,\ \tilde{\xi}=-\frac{\left(n-1\right)!!}
{\left(n-2\right)!!}\frac{1}{\sqrt{2^n}}\tilde{\mu},$ and
 $B_{2m}$ is the Bernoulli number.
The first term is the contribution from the planar singularity
 which can be read off from eq.~(\ref{eq:rhop}).
  We see that the form of the higher genus contributions
  as a function of the scaled Fermi level is the same as the
 one at the lowest critical point.

  Let us finally point out that  this result implies a strange scaling
 behavior not suggested by the standard continuum analysis.
  The final answer for the higher genus contribution as the renormalized
  cosmological constant $t$ is obtained  by
 substituting eq.~(\ref{eq:kx}) into   eq.~(\ref{eq:rhoas}):
 $ \tilde{\mu} = N \mu \sim N x^{\frac{2n}{n+1}}$.
  As $x$ goes like $a^{2}t$, we conclude that the renormalized string coupling
 $\kappa$   must be defined as
\beq
\label{eq:sc}
 \kappa \sim \frac{1}{N a^{\frac{4n}{n+1}} }\;\;\;.
\eeq
 To make the planar ground state energy finite, on the other hand,
  one has to have  a formula
\beq
  \kappa^{(0)} \sim \frac{1}{N a^{2- \gamma_{str}} } =
 \frac{1}{ N a^{2 + \frac{n-1}{n+1}} } \;\;,
\eeq
  which is in agreement with
 the standard formula  for $c\leq 1$
   but is not in agreement with  eq.~(\ref{eq:sc}).
 Let us adopt eq.~(\ref{eq:sc}).  We either have to subtract the planar
  contribution  to obtain
\beqn
 E_{gr} \sim \frac{2^{\frac{3}{2}n-1} }{4 \pi} \frac{(n-2)!!}{(n-1)!!}
 \sum_{m=1}^{\infty} ( 2^{2m+1} - 1 ) \frac{\mid B_{2m+2} \mid}{(m+1)m}
  \kappa^{2m} \left(ct^{\frac{2n}{n+1}} \right)^{-2m} \;\;\;,
\eeqn
  with $c$ an  $n$ dependent constant,
or to multiply the complete ground state energy by $a^{2-\frac{4}{n+1}}$
 to  take the continuum limit,
 in which case we get a trivial result  $E_{gr} \sim \frac{1}{\kappa^{2}}$.
 This disparity in the scaling behavior
 between the planar result and the one from the higher genera
 seems to suggest a connection of the multicritical points  with
  discrete states which appear only in the external lines\cite{GKN,Pol}.

\end{document}